\documentclass[usenatbib]{mn2e}
\usepackage{graphicx}

\begin{document}
\topmargin-1cm

%Make my life significantly easier
\newcommand\approxgt{\mbox{$^{>}\hspace{-0.24cm}_{\sim}$}}
\newcommand\approxlt{\mbox{$^{<}\hspace{-0.24cm}_{\sim}$}}
\newcommand{\be}{\begin{equation}}
\newcommand{\ee}{\end{equation}}
\newcommand{\bea}{\begin{eqnarray}}
\newcommand{\eea}{\end{eqnarray}}
\newcommand{\mmin}{M_{\rm{min}}}

\title{Issues in joint SZ and optical cluster finding}
\author[Cohn \& White]{J.D. Cohn${}^{1}$ and Martin White${}^{2}$\\
${}^1$Space Sciences Laboratory,\\ 
${}^2$Departments of Physics and Astronomy,\\
University of California, Berkeley, CA 94720}

\date{\today}
\maketitle
\begin{abstract}
We apply simple optical and SZ cluster finders to mock galaxy catalogues and
SZ flux maps created from dark matter halos in a $(1\,h^{-1}$Gpc${})^3$ dark
matter simulation, at redshifts 0.5 and 0.9.
At each redshift, the two catalogues are then combined to assess how well
they can improve each other, and compared to several variants of catalogues
made using SZ flux and galaxy information simultaneously.
We use several different criteria to compare the catalogues, and illustrate
some of the tradeoffs which arise in tuning the galaxy cluster finders with
respect to these criteria.
We detail many of the resulting improvements and issues which arise in
comparing and combining these two types of data sets.
\end{abstract}

\section{Introduction}

In order to study galaxy clusters one must first find, and ideally weigh,
them.  Catalogues spanning a range of redshifts, with a well understood
selection function, are key to learning about many cluster properties --
for a recent review, see e.g.~\citet{Voi05}.
These include counting clusters as a function of redshift to study dark
energy and cosmological parameters, 
characterizing them as environments for galaxy formation, and for
understanding galaxy cluster formation and evolution itself.
As galaxy clusters correspond to the largest dark matter halos in numerical
simulations, many of the statistical properties of the galaxy cluster
population are theoretically accessible.  Indeed, predictions for dark matter
halo properties in collisionless N-body simulations are converging
\citep{Evr08,CodeCompare,Luk08,Tin08}.
The number and positions of cluster sized halos (their counts and clustering) 
are of interest for dark energy applications, while for cluster and galaxy
properties and evolution these halos comprise the backbone for the more
complex and less well understood cluster and galaxy gas physics.

In this note we study mock cluster samples produced by finding overdensities
of galaxies or distortions in the cosmic microwave background produced by the
upscattering of CMB photons by the hot intra-cluster gas
\citep[][hereafter SZ]{SunZel72,SunZel80}.
We compare catalogues found from both methods optimized separately, optimized 
as a pair, and using one catalogue as input for the second catalogue's search 
algorithm -- a first attempt at joint optical-SZ cluster finding.
We identify trends in the resulting samples, as well as new issues and
tradeoffs which arise once SZ and optical data are jointly available.
The general features we find can be expected to have bearing on surveys using
optical and SZ cluster finding together, such as the (optical)
Dark Energy Survey (DES\footnote{http://www.darkenergysurvey.org}) and the 
(SZ) South Pole Telescope (SPT\footnote{http://pole.uchicago.edu}) experiment.

The advantage of combining two cluster finding methods is that each
method has different strengths and weaknesses.  For example, SZ cluster
finders can successfully find high mass halos
(e.g.~$\geq 2 \times 10^{14} h^{-1} M_\odot$)
but do not give accurate redshifts\footnote{Though rough estimates are possible
using SZ morphology \citep{SchPfrZar05}.}, while optical cluster finders can
miscast a low mass halo as a high mass one, but are much better at
obtaining redshifts (albeit still with possible errors).
Both optical and SZ cluster finders suffer from projection effects but possibly
in different ways as they trace different properties of the baryonic matter.
(Projection has been an issue for optical cluster finding since the earliest
surveys, e.g.~\citet{Abe58,Dal92,Lum92,Whi99}, and has been estimated 
in SZ by \citet{WhiHerSpr02,HolMcCBab07,Hal07,ShaHolBod07}.)
As the largest objects which have had time to virialize in the universe,
``average'' cluster properties evolve with redshift, and so the results
of joining SZ and optical may evolve as well.

The outline of our paper is as follows.
We start by describing the simulations and (standard) catalogue/map
constructions in \S 2.  In section \S 3 we describe the (again
standard) optical and
SZ cluster finders separately, and the resulting catalogues.  
 As our interest is in general
properties, we restrict ourselves to relatively simple cluster finders. 
Along with the finders,
criteria for assessing the success of a finder are introduced and
applied to a fiducial optical catalogue, for comparison with the
catalogues produced with the other finders in the rest of the paper.  
In \S 4, we compare the optical and SZ cluster catalogues to
the dark matter halo catalogues, and then introduce a matching between the 
optical and SZ catalogues directly. We measure 
how each catalogue is improved by the additional information
from the other.  Since the matching is not always one-to-one (one optical
cluster to one SZ cluster), the augmented
SZ catalogue (with optical information)
has different information than the improved 
optical catalogue (with SZ information).  Lastly,
a joint finder based on optical galaxies and SZ flux is introduced,
and two other steps in complexity are added, and these cases are 
compared to the previously discussed ones.

We identify the trends 
in all of these catalogues as a function of finder methods and parameters
using our success criteria. More detail about redshift success (a crucial use of
the optical catalogue and/or galaxies for SZ) and whether the SZ and optical
information can be combined to veto outliers are found in \S 5.
\S 6 summarizes, discussing the many improvements due to combining
the finders, trends which are
expected to generalize beyond our simple finders and catalogues,
 and issues which have been raised.

\section{Simulations: Dark matter, Galaxy Catalogues and SZ Maps}

N-body simulations provide a means of making mock catalogues with halos
situated in their correct cosmological context, capturing the population
properties, their environmental dependences, and projection effects.
We use a $1024^3$ particle dark matter simulation in a periodic box of side
$1\,h^{-1}$Gpc run with the TreePM code described in \citet{TreePM}.
This code compares well with others in code comparison tests
\citep[e.g.][]{CodeCompare,Evr08}.  The cosmological parameters,
$\Omega_m=0.25$, $\Omega_\Lambda=0.75$, $h=0.72$, $\Omega_Bh^2=0.0224$,
$n=0.97$ and $\sigma_8=0.8$, are in accord with a wide array of cosmological
observations.  The particle mass is $6\times 10^{11}\,h^{-1}M_\odot$ and
the force softening length is $35\,h^{-1}$kpc, allowing us to easily resolve
the halos of interest.
Further details of the specific simulation can be found in \citet{Bro08}.
We use fixed time outputs at $z=0.5$ and $z=0.9$ and make use of the
periodicity of the simulation volume to avoid issues concerning boundary
effects in the maps or catalogs (otherwise the boxes would span about 0.4 (0.6) 
at $z=0.5 (0.9)$ in redshift).  Periodicity allows objects in
any part of the box to be subject to analogous projection effects.

For each output we identify dark matter halos in three dimensions using the
Friends-of-Friends \citep{DEFW} algorithm, with linking length $0.168$ times
the mean interparticle spacing\footnote{This defines halos to be the material
above a density threshold of roughly $\rho>3/(2\pi b^3)\simeq 100$ times the
background density, with no requirement on the halo shape.}.
This is slightly smaller than the oft used $0.2$ linking length.  We found
that the longer linking length could over-merge halos; with our choice we only
link two regions if the density in the bridge connecting them is
$>100\,\rho_b$ everywhere.  We shall use these FoF masses throughout.

At $z= 0.5$ there are 7173 (1516) halos with
$M\ge 1(2)\times 10^{14}\,h^{-1}M_\odot$, compared with 2167 (295) at $z=0.9$.
These halos will be the sought after objects in our cluster searches.  
The observables (red galaxies and SZ flux) 
of the halos are added to the simulation in
post processing as described below.

\subsection{Red Galaxy Mock catalogues}

Observationally cluster galaxies are predominantly red, with little or no
on-going star formation.  Most clusters at low $z$ are observed to have a
tight ``red sequence'', and overdensities of galaxies in this sequence can
be used to find clusters
\citep{BowLucEll92,LoC97,GlaYee00,LoCBarYee04,GalLubSqu05,GlaYee05,Gla06,
Wiletal06}.
In order to mimic these galaxies, we populate the halos in our simulation with
red galaxies in a manner which reproduces the luminosity function and
luminosity dependent clustering seen in the NDWFS \citep{NDWFS}, as modeled
in \citet{Whi07} and \citet{Bro08}, over a range of redshifts.
The clustering of our model galaxies is remarkably consistent with that found
subsequently for 2SLAQ \citep[][Ross, private communication]{Ros07}.

We model photometric redshift errors for these (red) galaxies as a Gaussian
of width $\delta z=0.02(1+z)$.
(We explored trends with larger and smaller photometric errors, which degraded
or improved the finders as expected.)  
We did not include a population of objects with `catastrophic' redshift
failures, or a population of blue galaxies for which photometric redshifts
are generally less accurate.  Models of these objects are less well developed.
As we will see below, the optical cluster finding was
difficult enough without adding these complications and we wanted to focus
on the main features of the method rather than the details which would need
to be addressed by a finder operating on observational data.

We assumed that the photo-$z$ errors were uncorrelated for different galaxies
within each cluster, unless stated otherwise.  If photo-$z$ errors were
correlated for galaxies within the same cluster -- perhaps due to shared
evolution -- the observed cluster would shift rather than spread in the
redshift direction.  We looked at 10 clusters in the SDSS C4 \citep{Mil05}
catalogue lying in the equatorial stripe in the range $0.07<z<0.15$, for which
both spectroscopic and photometric redshifts of the member galaxies are known.
For these it appears that the photometric redshift errors are at most weakly
correlated.  The upside of this is that the error on the mean redshift of the
cluster galaxies is significantly tighter than on an individual galaxy.
While we cannot be sure that this decorrelation holds at higher redshift or
with different filters, we note that some scatter in photo-$z$ is expected
{}from photometry errors in any case, so we can consider our approach to be
conservative for finding clusters.

The resulting mock galaxy catalogues have objects with a 3D position,
luminosity and an observed photo-$z$.  We keep all galaxies brighter than
$\frac{1}{2}L_*$, giving $\sim 4 (3)\times 10^6$ galaxies at redshifts 0.5
(0.9).

\subsection{SZ flux maps}

The dark matter simulations are also used to create SZ $y$-distortion maps for
both redshifts, using the method described in detail in
\citet{SchWhi03} and \citet{ValWhi06}.
To recap briefly, the gas is assumed to follow the dark matter density and
to be at a uniform temperature $T\propto M^{2/3}$, with a low-$M$ cut-off
for halos below $1\,$keV.
The $y$-distortion is computed by projecting through the material in the
box, using the non-relativistic expression \citep{SunZel72,SunZel80}.
These approximations are poor in detail, and can be easily improved upon, but
provide a good approximation to more realistic $y$ maps produced by
hydrodynamic simulations \citep[e.g.][]{WhiHerSpr02} when smoothed with the
$1'$ beam of upcoming experiments (e.g.~SPT).
Such a beam subtends $400\,h^{-1}$kpc at $z=0.5$ and $600\,h^{-1}$kpc at
$z=0.9$, making the maps largely insensitive to the detailed cluster physics
and to our N-body resolution.
Our cluster finding, described below, also uses the integrated flux across
the whole peak, as this is expected to be a more robust measure of cluster
mass.  This approach down-weights the omitted effects even more.  We note
however that by assigning $T$ given the halo's mass, and ignoring its
dynamical state, we are likely underestimating the scatter in the $Y-M$
relation.

Assuming the distant observer, or plane parallel, approximation, a map for
each redshift of
$4096^2$ pixels is created by projecting through the $1\,h^{-1}$Gpc length
of the box.  As for the galaxy catalogs, the resulting maps are periodic,
allowing us to ignore edge effects below.
The pixels subtend $0.6'$ at $z=0.5$ and $0.4'$ at $z=0.9$, comfortably below
the $1'$ instrument beam we shall assume below.
We converted from Compton $y$-parameter to temperature distortion assuming
an observing frequency of $\nu/56.84\,{\rm GHz}\ll 1$, though this choice
is degenerate with our assumed normalization of the $T-M$ relation.
The overall normalizations of the maps was set by comparing $T$ and $Y$ as a
function of halo mass in the simulation to the similar correlations observed
in X-ray by \citet{ArnPoiPra07}.

The resulting maps were smoothed with a Gaussian of width $1'$, to approximate
the instrumental response, and gaussian noise of $20\,\mu$K-arcmin was added.
The resolution and noise were chosen to be similar to that expected for SPT
\citep{SPT}.  This is a best case estimate, as it ignores issues of CMB
cleaning, foreground contamination or systematics.
In order to increase contrast of the peaks in our maps, we smoothed it once
again by a Gaussian of width $1'$ and it is these maps which we used for our
cluster finding.

\section{Optical and SZ Cluster Finding}

We start by constructing separate optical and SZ cluster catalogues, 
such as two
separate surveys might produce, using simple finders to focus on general
trends.  
We will call the found objects ``clusters'', while the true objects in the
underlying dark matter catalogue will be called ``halos''.  Our main focus
will be on ``massive halos'', by which we mean those with
$M\geq 2\times 10^{14}\,h^{-1}M_\odot$.
Massive halos in our cosmology have
$\bar{n}\simeq 1.6\times 10^{-6}\,(h^{-1}{\rm Mpc})^{-3}$ and 
$\sim 2.8\times 10^{-7}\,(h^{-1}{\rm Mpc})^{-3}$ at $z=0.5$ and $0.9$
respectively.

\subsection{Optical Finder}

Our  optical finder  is a refinement of the circular overdensity method
used in \citet{Coh07}.  
The galaxies are first sorted by luminosity.  Starting with the
most luminous each galaxy is considered in turn as a potential cluster
center.  Every other galaxy is given a weight dependent upon the galaxy's
estimated position from the candidate cluster center,
\begin{equation}
  g(\delta r_3)=
  \exp\left[-\frac{1}{18}\left(\frac{|\delta r_3|}{\sigma}\right)^3\right]
\end{equation}  
where $\sigma$ is the photometric redshift error, expressed in $h^{-1}\,$Mpc,
and $\delta r_3$ is the distance from the central galaxy in the redshift
direction. 
This weight $g$ is multiplied by a second weight depending upon previous
membership in other clusters\footnote{This means
that changing parameters to find different clusters is different than finding
a larger cluster sample and then making a cut on the full sample afterwards.
Each galaxy contributes to potential clusters based on its membership in
previous ones.}.  If we write $r_{\rm clus}$ for the radius
of the found cluster, and $r_\perp$ for the radial distance of the galaxy
in two dimensions from the cluster center then the second weight for
each galaxy starts as
unity and is decreased by
\begin{equation}
  \Xi(r_\perp,\delta r_3)=g(\delta r_3)
  \left(1+16 \frac{r_\perp^2}{r_{\rm clus}^2}\right)^{-1}
\end{equation}
each time the galaxy is included in a cluster with
$g(\delta r_3)>g(3\sigma)\simeq 0.223$.
The second weight for a galaxy 
is set to zero if its value goes negative.

For each candidate center, starting with the 8 galaxies closest to
(and including) the center in the plane of the sky, the weights are summed
to define an overdensity
\begin{equation}
  \Delta \equiv
  \frac{\sum_{i} g(\delta r_{3,i}) \Xi(r_{\perp,i},r_{3,i})}
       {\pi r^2 \bar{n}_{gal}} 
\end{equation}
where the index $i$ refers to the $i$th galaxy, $r$ is the distance to the
farthest of the 8 galaxies and $\bar{n}_{\rm gal}$ is the average surface
density of galaxies.  We calculated $\bar{n}_{\rm gal}$ by taking the average
around 1000 randomly chosen galaxies in the simulation.
The cluster is kept if $\Delta\geq\Delta_p$, a critical (or threshold) density 
and a parameter in our finder.  For any cluster found, the nearest galaxies
are added until $\Delta<\Delta_p$ and all contributing galaxies which have
$g(\delta r_3)> g(3\sigma)$ are removed from the list of subsequent potential
cluster centers and given a decreased weight $\Xi(r_\perp,r_3)$ as described
above. 

As usual with overdensity finders, choosing a high value of $\Delta_p$ picks
out the halo cores and misses many halos entirely, while a lower value blends
many halos into one cluster.  After some experimentation we set $\Delta_p$
by measuring its value on $\sim10^3$ massive halos
($M \geq 2\times 10^{14}\,h^{-1}M_\odot$)
using the radius containing 55\% of the halo galaxies.  Our fiducial threshold
(below)
is the $67^{\rm th}$ percentile of the $\Delta_p$ distribution of these
objects.

The halo whose galaxies contribute the most to the weighted sum over the
cluster galaxies we call the best match halo, and $f_{\rm best}$ is defined
to be the best matched halo's fractional contribution to the full cluster's
weight\footnote{This is analogous to the ``Largest Group Fraction'' and
``Largest Associated Group'' used by
\citet[e.g.][]{Ger05}.}
(Another possibility for matching to a halo is to take the halo of
the central galaxy in the cluster, the analogue of the BCG.  We found that
this doesn't work as well using our criteria for success, described below.)  
The mass of the best matched halo is assigned to the cluster.
Clusters with $f_{\rm best}\geq 0.5$ will be considered clean,
clusters with lower values of $f_{\rm best}$ will be called blends.

We define the cluster's richness to be the sum of its galaxy weights and we
only consider clusters with richness $\geq 10$ hereon.
The optical cluster finding thus produces a list of halos with the ``richest''
cluster associated to each and a list of clusters with a richness, a
redshift (of the central galaxy), a best matched halo (its mass and three
dimensional position) and a measure, $f_{\rm best}$, of how good of a match
this halo is.

\subsection{SZ finder}

Cluster finding in the SZ flux maps follows \citet{SchWhi03}.
Peaks above $4\sigma_{\rm noise}$ ($80\,\mu$K-arcmin${}^2$) in the map are
identified and clusters are grown around these until all contiguous pixels
with flux at least 25\% of the maximum are included.
In most of the discussion below (unless specified otherwise) adjacent peak
regions are merged and assigned to the highest peak.
Each patch is then associated with the sum of fluxes from each of its pixels.
The flux overdensities are matched to halos with centers within two pixels
($\sim 500\,h^{-1}$kpc) of the patch; the halo with the largest mass is taken
to be the best matched halo.
Most ($>95\%$) of the peaks were matched to $< 6$ halos of mass
$>10^{13}\,h^{-1}M_\odot$, the most matched peak had 18 matches.
The number of matches is due to the sky area subtended by each patch (on
average about 1 $(h^{-1} Mpc)^2$ or $\sim 18$ pixels)
and the lack of line-of-sight information from the SZ map itself.
The number of halos per patch is larger than expected based on purely Poisson
statistics (given the very low number densities of halos), but the halos
exhibit significant clustering, enhancing the number within any
given SZ patch.

The resulting SZ cluster catalogue contains the SZ flux and two dimensional
peak position for each patch, and the best matched (most massive) halo and
its three dimensional position.  We also kept the number of pixels in the
SZ patch but did not use it in the analysis as it is $\simeq$75\% correlated
with the total SZ flux in the patch.

\section{Optical and SZ Catalogue Comparison} \label{sec:optszcats}

We now have three catalogues in hand: the optical cluster catalogue, 
the SZ cluster catalogue and the halo catalogue (our ``truth'').
A perfect finder would find every massive halo as a cluster once and only
once, with the finder observables richness and flux that are well
correlated with its mass and -- for the optical catalogue -- a correct
redshift.  However, catalogues can succeed better or worse at different
combinations of these properties.  

In order to compare the catalogues, we measure the following
quantities, which all tend to zero in the ideal case:
\begin{itemize}
\item the number of halos with $M \geq 2\times 10^{14} h^{-1} M_\odot$ which
are not the best match halo of any cluster (missed)
\item the number of
blends (clusters with $f_{\rm best} < 0.5$),
for clusters with an optical richness only
\item the scatter in the least squares fit to the 
lg(richness)-lg(mass) relation (and lg(flux)-lg(mass) when SZ is 
included),
\item the number of overcounted massive halos
(the number of massive halos found as best match minus the number found 
as best match which
are unique, divided by the total number of massive halos in the box).
\end{itemize}
In addition, the separation between the measured and true redshift will be
measured for finders which provide a redshift.

These criteria are not all independent.
Blends, for example, will increase scatter in the richness-mass relation, but
also tend to change the shape of the scatter \citep{Coh07}.  The mass
distribution for blends, for a given richness cut, is centered around a
lower mass than that for the clean clusters.  One
might want to favor one criterion strongly over another however.
For instance, if one is building a catalogue for X-ray followup,
rather than measuring the statistical properties of the
full halo distribution, one might want to minimize blends to 
the extreme at the cost of losing many massive halos.
 
\subsection{Optical Cluster results}

\begin{figure}
\begin{center}
\resizebox{3.5in}{!}{\includegraphics{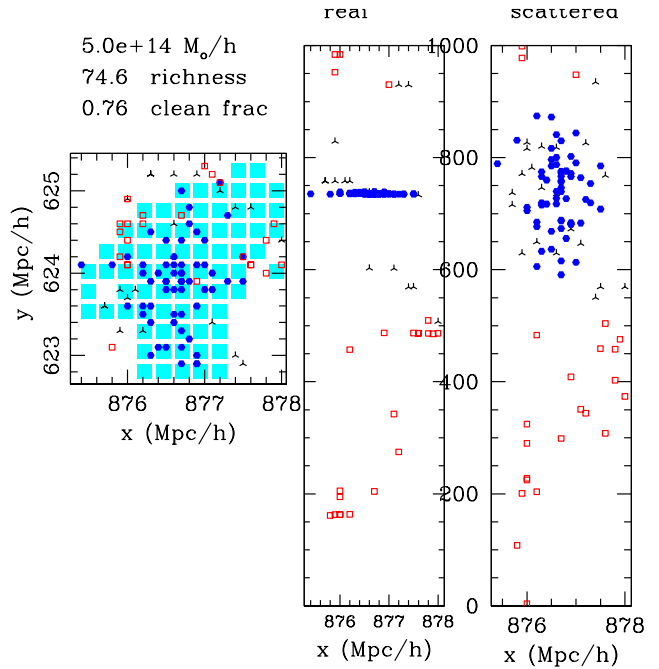}}
\end{center}
\vspace{-0.1in}
\caption{A rich clean cluster.
At left is the view on the plane of the sky.  The shaded regions are
pixels of SZ flux in the corresponding found SZ cluster.  
The filled circles are galaxies
belonging to the halo contributing the most richness to the cluster, the
little ``T'' marks are galaxies not from the dominant halo, but still
contributing richness $\geq 1/3$ to the cluster, and the open squares are
galaxies who each contribute richness less than 1/3 (and thus are not
really considered part of the cluster).  At right, the two long stripes show
the position of the galaxies in the
x vs. redshift direction. The scale in the redshift (vertical) direction runs 
the full $1000\,h^{-1}$Mpc of the box.  The true redshift direction
positions are shown in the lefthand stripe, and 
the observed, scattered positions are
shown in the righthand stripe.  The ``puffing out'' of the cluster in
the redshift direction is clearly visible.
}
\label{fig:observg}
\end{figure}

\begin{figure}
\begin{center}
\resizebox{3.5in}{!}{\includegraphics{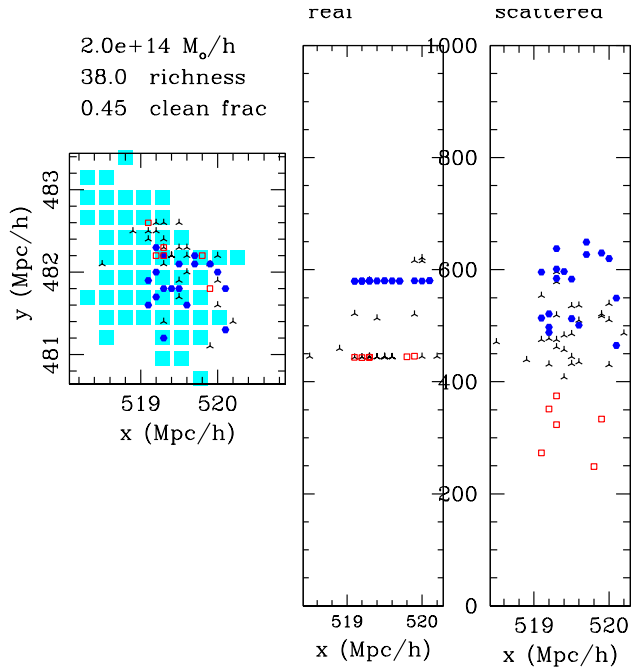}}
\end{center}
\vspace{-0.1in}
\caption{Example of a very rich blended cluster, with points having the same
meaning as in the previous figure.  Notice how the photo-$z$ errors have
``puffed'' out the galaxy positions of two halos separated by
$>100\,h^{-1}$Mpc, mixing them up.
}
\label{fig:observb}
\end{figure}

To illustrate the range of behaviours we see in the simulations, we first
present some case studies.  Fig.~\ref{fig:observg} shows a very rich and
massive ($M=5\times 10^{14} h^{-1} M_\odot$) clean cluster.  The effect of
photo-$z$ scatter can be clearly seen in the ``puffing out'' of the galaxies
in the redshift direction, but the cluster is still identified by our finder.
Fig.~\ref{fig:observb} shows a more complex example.
This blended optical cluster has a $2\times 10^{14}\,h^{-1}M_\odot$ halo 
contributing most of its richness, but is blended with another halo of mass
$1.4\times 10^{14}\,h^{-1}M_\odot$ about $150\,h^{-1}$Mpc away.
In general the combination of photo-$z$ scatter and the non-circular halo
shape complicates the matching clusters and halos.
Halos can also be missed because they are abnormally poor for their mass
(the number of satellites in each halo is Poisson distributed) or because they
are too close to another halo and get confused with it.

\begin{figure*}
\begin{center}
\resizebox{6in}{!}{\includegraphics{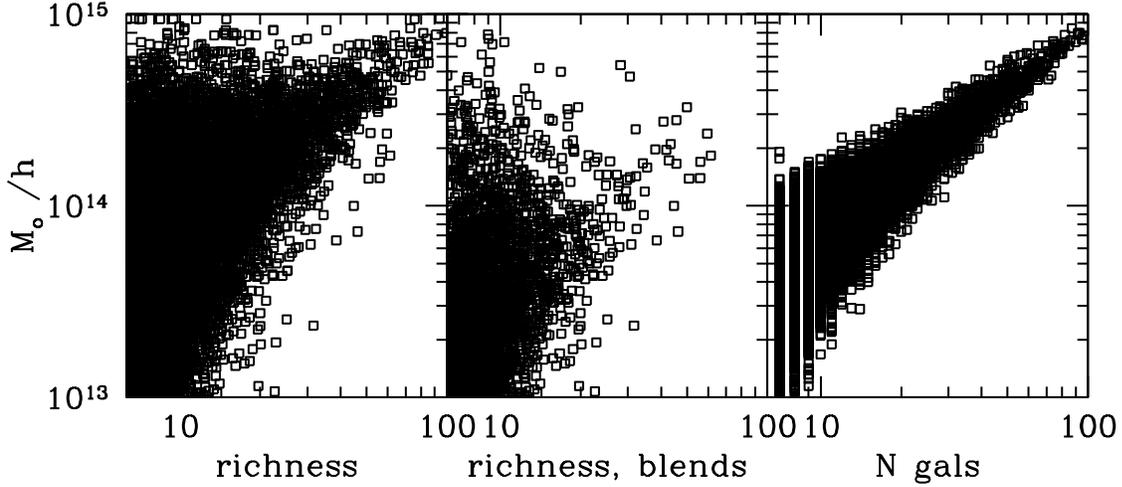}}
\end{center}
\vspace{-0.1in}
\caption{Left: the lg(richness)-lg(mass) relation for found clusters for
the fiducial model described in the text.
Center: the lg(richness)-lg(mass) relation for blends, those clusters with
less than half their richness coming from galaxies belonging to any one dark
matter halo.
Right:  the number of galaxies-mass relation for the dark matter halos.
The scatter for the found clusters is 0.30 in lg(mass), for those with 
richness $\geq10$, while for the halos it is 0.13 in lg(mass).
}
\label{fig:nofm}
\end{figure*}

For our fiducial choice of cluster finding parameters the optical catalogue
misses $\sim 15\%$ of the massive halos, has 14\% blends, a scatter in
lg(richness)-lg(mass) of 0.30 and the overcounts equal
17\% of the total number of massive halos.  (To clarify what we mean by
overcounts: 264 of the best match halos were already the best match halo
at least once.)  
The lg(richness)-lg(mass) relation is shown in Fig.\ref{fig:nofm}.
Recall that we require a minimum richness of 10 for a cluster to be included
in the optical catalogue.  This scatter is large, but note that there is
an intrinsic scatter in the lg($\#$ galaxies)-lg(mass) relation for the halos
of 0.13 lg (mass)\footnote{The lowest mass halos dominate our counts and have $10$
galaxies.  Ignoring the central and assuming Poisson statistics \citep{Koc03},
we expect $9\pm\sqrt{9}$ satellites at fixed mass or, since the $N-M$ relation
is close to linear, a $30\%$ scatter in $M$ at fixed $N$.  Since
lg$M\simeq 0.4\ln M$ this is $\delta\lg M\simeq 0.12$.}.
The richness-mass scatter can be reduced slightly by using 
the sum of cluster galaxy luminosities (for cluster galaxies with
$g(\delta r_3) >  g(3\sigma)$) as a mass indicator, and more
refined measures could possibly improve this further.  
As in \citet{Coh07}, for a given fixed richness cut the blends seem centered
at a lower mass than the clean clusters.  The blends thus produce a bimodality
in the richness-mass relation.
At redshift $z=0.9$, the catalogue has 14\% missed massive halos, 23\% blends,
$0.34$ scatter in lg(mass) and overcounts equalling 15\% of the massive
halos in number.  The increase
in blends with redshift occurred also with the color based finder in
\citet{Coh07}.

If instead of using the most contributing halo as the cluster halo, one takes
the halo of the central galaxy as the optical cluster halo
(``the BCG halo'')\footnote{We thank A. Leauthaud and R. Nichol for suggesting
we quote this number.}, the halo matching changes for 8\% of the $z=0.5$
optical clusters. 
The number of missed halos goes up as clusters more often get identified with
lower mass halos rather than the more massive ones.
Not surprisingly, the number of blends goes up as well: if the BCG halo and
most contributing halo coincide then the BCG halo is already the fiducial
optical cluster halo. 
A change to any other halo means a change to a halo contributing fewer galaxies
and thus a change to a blend.
The lg(richness)-lg(mass) scatter goes up (because more optical clusters are 
now matched with low mass halos) and the amount of overcounting goes down
slightly (from 17\% to 13\% of the number of massive halos).
This supports the use of the most contributing halo as the best match halo for
the optical cluster.  What appears to be happening at least some of the time
is that a massive halo is ``caught'' by a luminous galaxy which scatters into
its redshift range and becomes its center due to photo-$z$ errors.

Several tunable parameters/functions are part of the optical cluster finding:
the minimum density to be labelled a cluster, $\Delta_p$, the minimum number 
of galaxies to require before imposing the overdensity requirement, the 
weights in the redshift direction, and downweighting in radial direction.
We varied these to try to minimize the number of blends and maximize
the number of found massive halos.  For example,
changing the overdensity $\Delta_p$ can decrease the number of blends,
the scatter in the richness-mass relation and overcounting, but 
increases the number of missed halos, and vice versa.
We experimented with the weight $g$ in order to catch some of the 32\% of
cluster galaxies with more than the $1\sigma$ photo-$z$ errors, but not to
catch too many galaxies too far out.

Smaller observational errors can of course improve the cluster finding.
For example, in the extreme limit that the redshift errors are reduced to
$10\,h^{-1}$Mpc, if we again use the radius containing 55\% of the halo
galaxies to calculate 
$\Delta_p$, $\Delta_p$ goes up by a factor of 6, the finder misses 30\% more of
 the massive halos, cuts
the blends by 70\%, the scatter by 25\% and slightly reduces the 
overcounting.\footnote{Changing the redshift errors changes the contrast
with the background density for the clusters used to calibrate $\Delta_p$.}
Lowering $\Delta_p$ to miss fewer massive halos than the fiducial case
still decreases the blends by 70\%, the scatter by 20\% and the overcounting
(though not by as much).  In contrast, increasing the photo-z errors to 5\% and
changing $\Delta_p$ to miss the same number of massive halos as in the
fiducial case raises the number of blends by a factor of 2.4.

Another possibility, mentioned above, is that the photo-$z$ errors are
correlated within a given halo, due perhaps to similar star formation
histories for the member galaxies.  As noted we did not see such a trend
in the low-$z$ C4 catalog, but if present it would reduce the ``puffing
out'' in the redshift direction and make halos easier to find.
In the simulations we experimented with 
a fully correlated scatter in the redshift
direction and saw a strong improvement, as expected.
When all photo-$z$ errors are the same for galaxies within the same halo then
in comparison to the fiducial case the number missed is slightly smaller,
the number of blends drops by almost 50\%, the scatter by 10\%
and the overcounting by about 20\%.

The finder also does much better for the
$\sigma_8 = 0.9$ catalogues of \citet{Coh07}, which implies a non-trivial
dependence upon $\sigma_8$ of the finder at fixed mass cut.
Essentially, a lower $\sigma_8$ is analogous to fixing $\sigma_8$ and raising
the mass or redshift -- increasing the number of disturbed halos, increasing
the relative number of possible low mass interlopers for a given high mass halo
and decreasing the halos' contrast with the background more generally.
This has been seen in other studies: \citet[e.g.][]{Luk08} find that FoF and
SO finders differ more at high redshift and lower $\sigma_8$ universes.
This is very relevant to our situation as we are using an SO cluster finder
to search for FoF dark matter halos.

\subsection{SZ cluster results}

\begin{figure}
\begin{center}
\resizebox{3.5in}{!}{\includegraphics{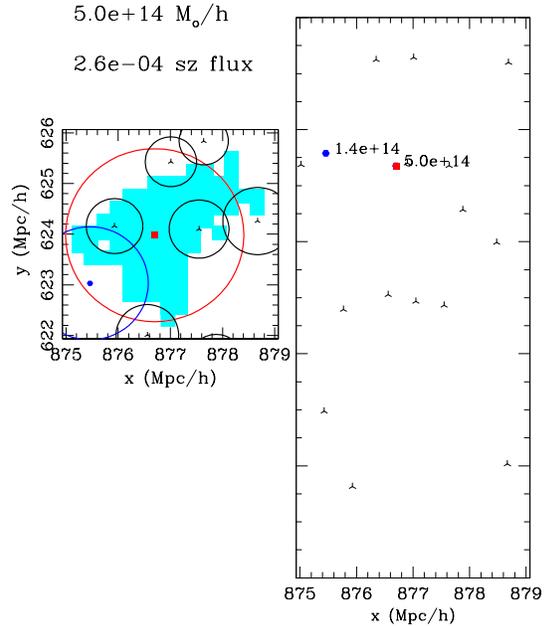}}
\end{center}
\vspace{-0.1in}
\caption{The left square is the SZ flux patch seen on the sky, and the filled
solid square is the center of the most massive halo identified with this patch,
corresponding to the cluster shown in Fig.~\ref{fig:observg}.
The filled dot at left is another $M \geq 10^{14} h^{-1} M_\odot$ halo
near the SZ cluster, and the little ``T'' symbols are centers of
$M\geq 10^{13} h^{-1} M_\odot$ halos in the region.  Circles are taken
to be $1\,h^{-1}$Mpc for the $10^{14}\,h^{-1}M_\odot$ halo, and scaled
by $M^{1/3}$ for other masses.  At right, the long axis is in the redshift
direction and goes the entire $1000\,h^{-1}$Mpc of the box, and shows the
halo positions in this
direction (indicating masses of $\geq 10^{14}h^{-1}M_\odot$ halos).
}
\label{fig:szexamp}
\end{figure}

We now turn to the SZ cluster finder, again starting with an example.
An example of an SZ cluster is shown in Fig.~\ref{fig:szexamp},
corresponding to the optical cluster shown earlier in
Fig.~\ref{fig:observg}.  The SZ cluster is centered on the most massive 
$5 \times 10^{14}h^{-1} M_\odot$ halo it contains, but also has a
$1.4 \times 10^{14} h^{-1} M_\odot$ halo matched to it.

\begin{figure}
\begin{center}
\resizebox{3.5in}{!}{\includegraphics{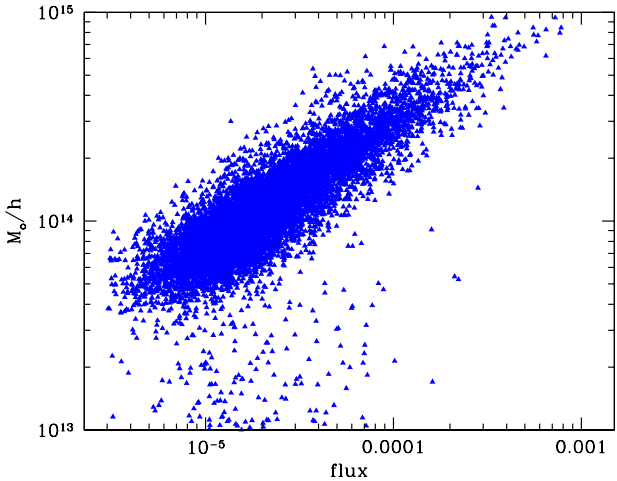}}
\end{center}
\vspace{-0.1in}
\caption{Mass for found SZ clusters at $z=0.5$ as a function of SZ flux
($\mu K\,$sr or $0.23\,T_{CMB}$arcmin${}^2$).}
\label{fig:SZfluxmass}
\end{figure}

The lg(flux)-lg(mass) relation for the SZ cluster catalogue on its own 
is shown in Fig.~\ref{fig:SZfluxmass}. % mofy sz_yfil_0.50.dat_1.0_szf_vsum_54.5
More than 98\% of the massive halos are found at $z=0.5$
(and $>99\%$ at $z=0.9$),
the lg(flux)-lg(mass) relation has a scatter of $\sim0.10$ in lg(mass) and
the overcounting is slightly smaller than the optical fiducial case ($13\%$).
For redshift $z=0.9$ the scatter is similar and the overcounting is $15\%$,
still smaller than fiducial optical value 17\%.
We do not apply the blend criterion to the SZ only clusters.

Almost all the massive halos which are missed are inside an SZ cluster,
but are not the most massive halo in that SZ cluster.
The tight relation between SZ flux and mass is known, and
seen as well in our simulations:
the scatter in lg(flux)-lg(M) is smaller than between lg(richness)-lg(M).
The scatter in flux in our simulations is caused by projection and is also
well known \citep{WhiHerSpr02,HolMcCBab07,Hal07,ShaHolBod07} -- see
\citet{HolMcCBab07} for a calculation of the dependence upon $\sigma_8$.
There are, as noted above, approximations in the maps which might
cause improvements in the SZ cluster finding.
The projection noise in our simulations might be slightly smaller
than in reality because we project only through the $1\,h^{-1}$Gpc box rather
than through the full path from $z\sim 10^3$ to $z\simeq 0$.
Our maps have also minimized the intrinsic dispersion in the flux--mass
relation because we assume a simple functional form for $T_{\rm gas}$ given
$M$, independent of the halo's history or other properties.
Finally we have optimistically assumed that foregrounds can be efficiently
removed, perhaps by multi-frequency observation.
With these caveats however we see that the SZ effect provides a very good
means of catching all clusters.  It also catches anything else that is hot
enough along the line of sight between $z\sim 10^3$ and $z\simeq 0$.

Using SZ flux alone does not give redshifts for the found halos -- although
a method to estimate rough redshifts has been proposed \citep{SchPfrZar05}.
Without redshifts one can still, for example, measure angular clustering and
counts of galaxy clusters for cosmological parameters
\citep[e.g.][]{Dia03,MeiBar03,MeiBar04,CohKad05}
but this has much less information than the 3-dimensional 
counterparts (e.g. \citet{ViaLid96,ViaLid99,HaiMohHol01,HolHaiMoh01,CarHolRee02,LevSchWhi02,WelBatKne02,BatWel03,WelBat03,MajMoh04,Moh05}, see the review by \citet{WelBat03} for more detailed references).

\section{Catalogue combinations} \label{sec:combinecats}

Given the three catalogues, each of which has objects with
some measure of their mass, comparisons and combinations can now be made.  
Most matchings tend to be many to one -- for example each dark matter halo
can be matched to multiple optical or SZ clusters.
We took each catalogue and found the best match in the other
two catalogues for each object.  
For the dark matter halos the added information is the SZ patch with
the most flux and the richest optical cluster.  The results of
this matching, in particular, the numbers of missed and
overcounted halos, were reported above.
The SZ cluster finder is much better at having at least one match for 
any massive halo in its catalogue, compared to the optical finder. (The
$z=0.5$ optical finder finds 2 of the 36 halos missed by the fiducial SZ
finder, the fiducial SZ finder finds 193 of the 227 missed by the
optical finder.) 

To associate found optical and SZ clusters to each other directly, without
going through dark matter halos, we used a direct analogue of the assignment
of halos to SZ patches.  This procedure could also be used with
observational data, where the true dark matter halos are not known.
All optically found clusters with centers within 2 or fewer pixels of the SZ
patch are assigned to the SZ patch (or cluster).
The optical match of the SZ cluster is the richest of these optical clusters.
If an optical cluster matches to more than one SZ cluster, it is assigned the
SZ cluster with the largest flux.  Generalizations of this are clear, but for
simplicity we started with only one match per optical or SZ cluster, so that
every optically found cluster gets an SZ flux (which might be zero) and every
SZ found cluster gets an optical richness (which again might be zero).
This matching does not require the corresponding optical and SZ halos to
agree.
For all optical clusters with richness $\geq 10$ and non-zero SZ flux,
$\sim8\%$ of the optical and SZ halos differ in the fiducial case,
with only a slight
increase in catalogues with lower $\Delta_p$.  

To get a more complete combined SZ and optical catalogue it is useful to
start with an optical catalogue which finds more massive halos than
the fiducial optical catalogue does.
The latter misses 15\% of the most massive halos when optimized on its own.
The idea is that an optical catalogue with more blends (lower $\Delta_p$)
can be combined with SZ information to cut out many these additional blends,
ultimately producing a catalogue with more of the massive halos.
As will be seen below, this does indeed work well in our simulations.

\subsection{Optical Catalogue Plus SZ Flux}

\begin{figure}
\begin{center}
\resizebox{3.5in}{!}{\includegraphics{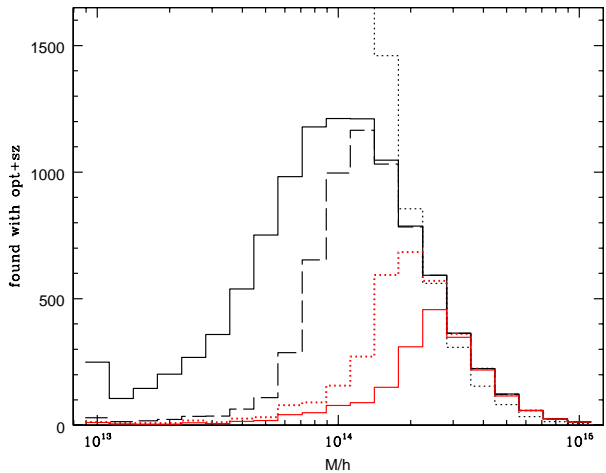}}
\end{center}
\vspace{-0.1in}
\caption{Number of massive halos found in the optical catalogue, requiring
flux $\Delta T\geq (0,1,3,5)\times 10^{-5}\,\mu$K sr
or $\Delta T\geq 0, 4.3, 13,21\times 10^{-5}\,T_{CMB}$arcmin${}^2$, from
left
to right (solid, dashed, dotted, solid).  The far left (solid) line 
is the mass distribution of the optical clusters with richness $\geq 10$ 
without any SZ cut, requiring the cluster to have nonzero SZ flux 
gives a line very close to the dashed line, second from left.
The high and lighter dotted line is the true number of massive halos,
overcounted by the finders at the high mass end. 
A threshold on
SZ flux is extremely good at weeding out low mass optical clusters.
}
\label{fig:optsz}
\end{figure}

The augmented optical catalogue is taken to include all optical clusters with
richness $\geq 10$ and a new requirement of SZ integrated flux above some 
cut:\footnote{The peak must also be above $4\sigma_{\rm noise}$.}
$\Delta T\geq 0, 1, 3, 5\times 10^{-5}\,\mu$K sr or
$\Delta T\geq 0, 4.3, 13,21\times 10^{-5}\,T_{CMB}$arcmin${}^2$ -- for
observations at frequency $\nu\ll 56.84\,$GHz this corresponds to an
integrated $Y$ parameter smaller by a factor of 2.
The number of massive halos found, as a function of mass and these cuts,
is shown in Fig.~\ref{fig:optsz}.

For the three SZ cuts, the three rightmost solid lines in
Fig.~\ref{fig:optsz}, the blend fraction goes down 25\%  from the fiducial
optical case, even though the optical catalogue alone has a relaxed overdensity
cut.  The lg(richness)-lg(M) scatter goes down by 33\%. The overcounting
increases, running from 24\%-20\% from left to right.  
The three cuts using SZ flux differ in the number of massive halos
missed.  From left to right the number missed is down from the fiducial
optical case by 40\%, then is only slightly fewer and for the
strongest cut misses almost twice as many.

\subsection{SZ Catalog Plus Optical Richness} \label{sec:SZopt}

In the previous section, each optical cluster was assigned an SZ flux
using the optical-SZ matching described above.
One can instead start with an SZ catalogue and add optical information, to
obtain an enhanced SZ catalogue.  (These are not identical because the
optical-SZ cluster matching is not one to one.)
In this case we take the SZ catalogue
with a cut of $3\times 10^{-5}\,\mu$K sr, second from right in
Fig.~\ref{fig:optsz}: the number of missed halos is very slightly
(from 15\% to 16\%) larger than in the fiducial optical catalogue,
but the blends and scatter in the lg(richness)-lg(mass) relation go down
by a factor of 2, 
the scatter in the lg(SZ flux)-lg(mass) relation goes up by $30\%$
and the overcounting goes down by 20\%.  If overcounting is a concern,
it is far better to start with the SZ catalogue and add optical richness
information.

\subsection{Outliers}

With observational data, one only has the optical and SZ cluster catalogues,
and not the underlying halo catalogues.  One question which arises is how
well the catalogues could be used together to flag outliers in the richness-
or flux-mass relations, i.e.~clusters which have an incorrect mass assignment.
The optical and SZ scatter of the ``true'' halo mass from that predicted by
the lg(flux)-lg(mass) or lg(richness)-lg(mass) relation are $\sim$50\%-70\%
correlated, depending upon how the catalogues are chosen.
This is not surprising as both SZ flux and galaxy richness suffer from
projection effects, so the scatter in mass is due in part to the same source. 
Even with this correlation, however, outliers can be reduced to some extent.

\begin{figure}
\begin{center}
\resizebox{3.5in}{!}{\includegraphics{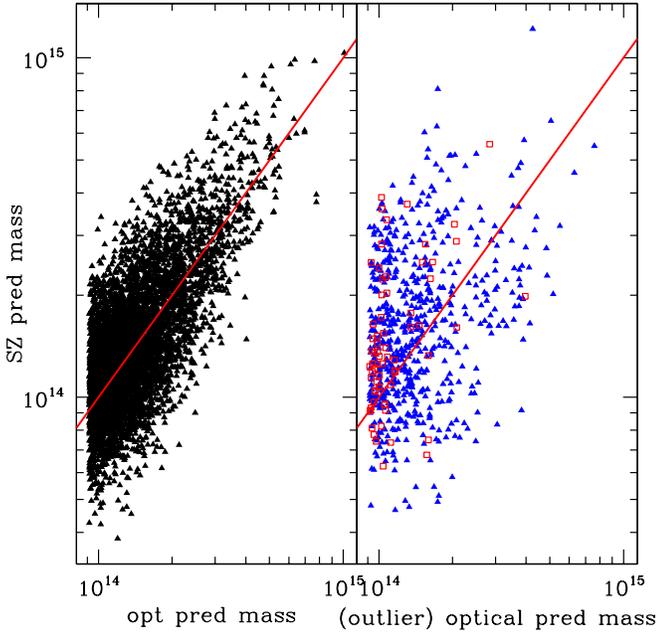}}
\end{center}
\vspace{-0.1in}
\caption{
Predicted mass using optical ``richness'' (x-axis) and 
SZ flux (y-axis) for clusters which are within 2 $\sigma$ of
the mean lg(flux)- or lg(richness)-lg(mass) relation
for both SZ and optical (left figure)
or above 2 (triangles) or 5$\sigma$ (squares) for either (right figure).
The halo mass, compared to the predicted optical or SZ mass,
is taken to be that corresponding to the optical cluster.
}
\label{fig:szoptscat}
\end{figure}

\begin{figure}
\begin{center}
\resizebox{3.5in}{!}{\includegraphics{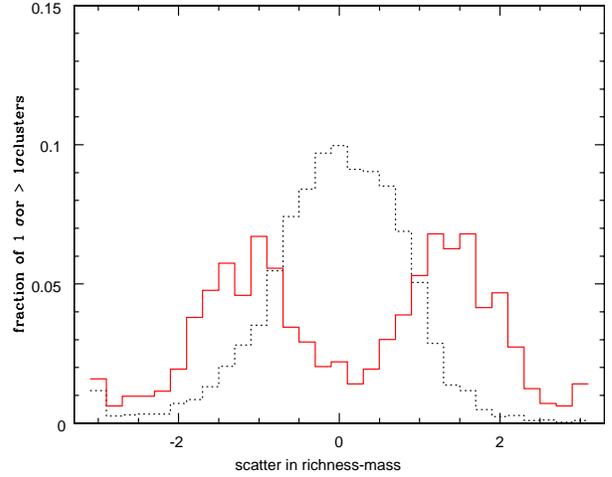}}
\end{center}
\vspace{-0.1in}
\caption{
Scatter off of optical lg(richness)-lg(mass) relation for
clusters which are within 1-$\sigma$ of the optical-SZ  predicted
mass relation in Fig.~\ref{fig:szoptscat} above (dotted line)
and those which are $\geq 1 \sigma$ off (solid line).
This suggests that making a 1-$\sigma$ cut on clusters using the
optical-mass vs. SZ mass relation will tend to cut out clusters which are 
outliers
in the lg(richness)-lg(mass) relation.
}
\label{fig:cmscat}
\end{figure}

Fig.~\ref{fig:szoptscat} shows where found clusters fall on the SZ and
optical mass relation if they are assigned both -- in this case we are
taking the true mass to be the mass of the best matched optical halo
for both the SZ and optical finders. 
At left are those clusters ($\sim 90\%$) which are $\leq 2 \sigma$ of the 
lg(richness)- or lg(flux)-lg(mass) relation, at right are those
further away from the mean relations.  Slightly over half of these $>2\sigma$
outliers in predicted to true mass are clusters which have
their SZ best matched halo not agreeing with their optical best matched
halo, and this is true of all of the $>5 \sigma$ outliers.
A discouraging trend to notice is that there are many clusters whose
SZ and optical predicted masses agree very well, which might lend confidence to
their values, but whose true masses are actually lying far
off (sometimes $> 5 \sigma$) of this relation. 
However, one can also take this relation and find the mean relation
(lg(flux) as a function of lg(richness) or vice versa) and cut out 
objects which lie more than $1\sigma$ off of it.
It appears most promising (Fig.~\ref{fig:cmscat}) 
to use SZ predicted mass as a function of optical predicted mass and 
cut $1 \sigma$ optical outliers on this relation -- these clusters seem
more likely to be true outliers in their predicted mass as a function of
optical richness.

\begin{figure}
\begin{center}
\resizebox{3.5in}{!}{\includegraphics{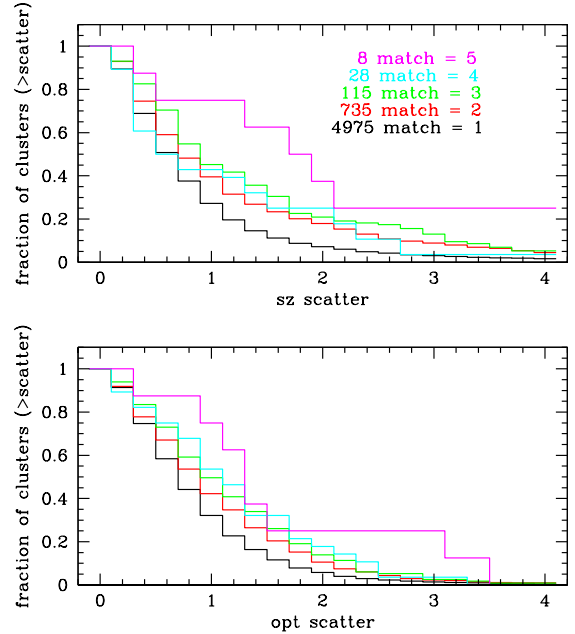}}
\end{center}
\vspace{-0.1in}
\caption{
The fraction of clusters with scatter above a certain amount in units 
of the width $\sigma$ of the mean lg(flux) or 
lg(richness)-lg(mass) relation).  
SZ clusters with more optical cluster matches tend to
have a larger scatter off of the mean relation, both for the SZ predicted mass
and the optical predicted mass.  The lines go from most to fewest matches
of optical to SZ patches, essentially from top to bottom, in each picture.
}
\label{fig:matchscatter}
\end{figure}

There is also a small correlation with number of optical clusters in an SZ 
patch and the size of the scatter of the SZ or its best matched optical 
cluster's mass from
the mean relation\footnote{We thank A. Leauthaud for suggesting we 
plot this quantity.} as can be seen in Fig.~\ref{fig:matchscatter}.
That is, SZ clusters with more optical cluster matches are more likely
to be off of the mean lg(flux) or lg(richness)-lg(mass) relation.

\section{Joint SZ-optical cluster finding} \label{sec:SZcenopt}

Instead of finding clusters separately using galaxy and SZ flux
maps and then combining catalogues one can use the optical and SZ
information concurrently to find clusters.  We investigated a number
of simple ways of doing this joint cluster finding.

A straightforward generalization of the optical finder is to start with
the SZ peak catalogue and then use each peak as a potential center for
optical cluster finding.
The optical finding is identical to that above (centered on a given
luminous galaxy) except that there is no a priori redshift for the candidate 
cluster center.  Instead, the finder scans through 50 different redshift bins,
each of width $20\,h^{-1}$Mpc, and keeps the cluster with the most richness
for each SZ peak.
Again the overdensity $\Delta_p$ is tuned to the most massive halos,
calibrated in this case for the most massive halos in the dark matter
simulation using an offset of the diagonal length across a pixel
($\sim 0.35\,h^{-1}$Mpc) as the SZ peak position can be anywhere within
the pixel.

This joint finding is a marked improvement compared to the cases
above (optical alone, or merging separate SZ and optical catalogues).
Although the number of missed massive 
halos remains about the same, the number of blends goes down
by almost a factor of 2, the lg(richness)-lg(mass) scatter goes down
by $>40\%$, the lg(flux)-lg(mass) scatter is about the same,
and the overcounting goes down from a factor of 20\% to 2\%.
Essentially, by using the SZ peak as a way to find cluster centers, one
is more likely to start near a cluster center.

Two extensions were explored as well.
As many halos were ``lost'' because they weren't the dominant halo in
an SZ patch, one can use the SZ patch as an indicator of hot gas
and then search for not one, but many clusters within each SZ patch.
After finding the first SZ peak-centered optical cluster, we sorted
the remaining galaxies in order of decreasing luminosity and 
repeated the usual optical cluster finding.  One finds that the more
such clusters one includes (e.g. keeping up
to 3 of the clusters in each peak),
the fewer rich halos are missed.  However,
the success with respect to the rest of the criteria worsen, 
compared to the 1 optical cluster
per SZ peak 
case above:  blends, scatters and overcounting all increase.

A second generalization is to start instead with the
unmerged SZ flux maps.  In this case, one is using all the local
SZ peaks as cluster finding centers, even if the peak is in a patch
that adjoins
another, rather than just the largest peak in the contiguous region.
Doing this for the SZ cluster finding alone increases the lg(flux)-lg(mass)
scatter by 40\%, and results in an overcounting by 
almost twice the number of massive halos in the box  
(80\%, $\sim 6$ times higher than the unmerged
SZ cluster finder).  
However, using unmerged SZ peaks as a starting point for optical
cluster finding allows the optical cluster search to cut down the overcounting.
For this case (one cluster per unmerged SZ peak),
the number of halos compared to the fiducial optical case which are
missed goes down by $\sim 33\%$, the blends go down by $\sim 33\%$, the
overcounting barely decreases and the scatter
in lg(richness)-lg(mass) goes down by $\sim 33\%$.
The scatter in lg(flux)-lg(mass) goes up by $\sim 30\%$
.

\section{General Trends}

\begin{figure}
\begin{center}
\resizebox{3.5in}{!}{\includegraphics{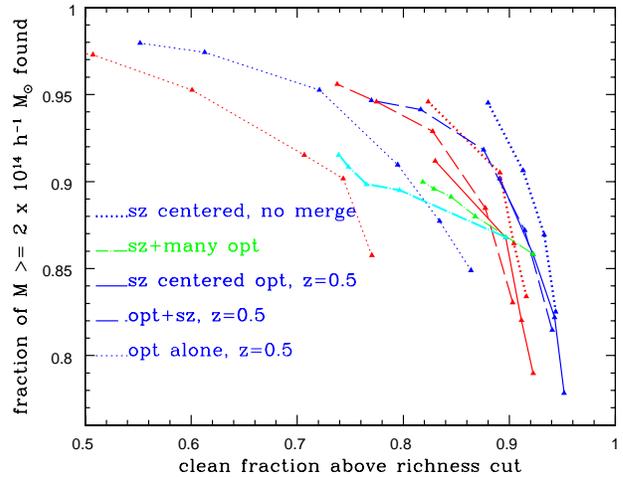}}
\end{center}
\vspace{-0.1in}
\caption{A comparison of clean fractions and completeness for several
different cluster finders.  The best finders are towards the upper
righthand corner, having a high clean fraction and a high fraction of found
massive halos.  The light solid dotted lines (upper $z=0.5$, lower $z=0.9$)
correspond to the optical search alone, changing $\Delta_p$.
Increasing $\Delta_p$ increases the clean fraction, but decreases the
fraction of massive halos found.
The dashed lines are the combined SZ and optical catalogues,
the solid lines are optical searches around SZ cluster peaks and
the heavy dotted lines are optical searches around SZ cluster peaks
where the SZ cluster map does not merge adjoining clusters,
again all for variation of the optical overdensity parameter.
Also shown are the results where optical clusters are searched for in SZ
clusters starting with a search around the central SZ peak position and
then continuing to look for other optical clusters within the same SZ cluster.
In this case, the number of additional clusters changes along the line,
starting with 1 and then increasing.  As more clusters per patch are included,
more halos are found, but many of them are blends;
the clean fraction decreases. 
}
\label{fig:comparem}
\end{figure}
%collectstatscg

Generally, combining optical and SZ catalogues works substantially better
on all fronts than either individually -- as hoped -- and the joint finding
offers further improvements.  
The tradeoff between decreasing blends and increasing completeness can be
seen in Fig.~\ref{fig:comparem}, as a function of changing overdensity cut
$\Delta_p$.
To summarize the different methods shown:
\begin{itemize}
\item Optical alone (light dotted lines at left): the fiducial optical model.
\item Optical + SZ (dashed lines): 
using an SZ catalogue and optical catalogue together, keeping
all SZ clusters
which have a matching optical cluster.  The optical catalogue has
a relaxed (lower) overdensity threshold compared to the fiducial optical
catalogue, to be more complete
(it correspondingly has more blends on its own than the fiducial optical catalogue).
\item SZ centered optical (solid lines): for each SZ peak, search for an optical cluster
using the SZ peak as the center position, taking the redshift to be
that for the center which gives the richest cluster.
\item SZ + many optical (heavy dot-dashed): 
for each SZ peak, search for an optical cluster using 
the SZ peak as the center, as above, then continue to look for more 
optical clusters using the rest of the galaxies in the SZ
patch as possible centers, in order of decreasing luminosity.
Moving along the line includes more and more of
these other optical clusters in the same patch, in the order they
were found.
\item SZ centered, no merging (heavy dotted): 
in the SZ finding, instead of merging all SZ clusters
whose regions touch into one cluster, with one main peak, take
each unmerged peak as a separate possible optical cluster center and search
around it.
\end{itemize}
In all cases, the upper line in a given pattern 
is for $z=0.5$; the combined 
finders all work better at lower redshift.  Ideally the cluster finders
would give results far to the right (large clean fraction) and far upwards 
(large fraction found), but the tradeoff between clean fraction
and fraction found (or purity vs. completeness) is evident in the
trajectories for each method.

From this diagram, one sees that the most clean clusters are obtained by using
either optical cluster finding around SZ cluster peaks (merged or unmerged)
or using one optical cluster match to each SZ peak by taking an SZ catalogue
and matching one optical cluster to each SZ cluster.
However these are not all equivalent in terms of the other criteria,
the numbers for which are listed in Tables \ref{tab:props5} and 
Tables \ref{tab:props9}.
In particular, although the scatters are similar for all three, the (merged)
SZ centered optical search has much less overcounting than the other two
methods.  (Recall this overcounting is not with respect to the number
of uniquely found massive halos but with respect to the total number of
massive halos.  To get the former requires multiplying by the total number
of massive halos divided by the number of uniquely found halos.)
For $z\sim 0.9$, the optical finder is worse and alone gives slightly more
overcounting relative to $z \sim 0.5$ for the same $\Delta_p$ cuts.
When SZ is brought in (except for the no merging case) the overcounting drops
slightly, the SZ scatter is smaller and the optical scatter drops as well.
The optical cluster finder is expected to do worse at high redshift because of
the increased number of mergers (which are difficult to find in the plane of
the sky using a circular overdensity finder, reducing the number of found
halos) and the increase in relative number of possible interlopers per high
mass halo (as the latter decreases much more sharply than the former with
increasing redshift).  The high mass halos are also more biased at high 
redshift.
\begin{table*}
\begin{center}
\begin{tabular}{|l|l|l|l|l|} \hline
\hline method &blend fraction&scatter &scatter&overcounting \\
&(finding 90\% massive) & lg(richness)-lg(mass)&lg(flux)-lg(mass)& \\ \hline
optical only &19\% & $\geq 0.30$& -- & 17\%-33\% \\  \hline
SZ catalogue \\
+ optical catalogue &11\% &  0.18-0.19& 0.12& 14\%-17\%\\ \hline
SZ peak centered\\ optical search &11\% &0.17-0.18& 0.10-0.11& 1\%-3\% \\ \hline
SZ peak centered\\ without merging&8\% &0.19&0.12-0.13& 11\%-23\% \\ \hline
SZ peak centered\\ many optical&18\% & 0.18-0.29&0.11-0.31& 2\%-5\% \\ \hline
SZ alone (merged)&-- & -- & 0.10 & 13\% \\ \hline
SZ alone (unmerged)&-- & -- & 0.14    &75\% \\ \hline
\hline
\end{tabular}
\end{center}
\caption{ Properties of finders for redshift $z \sim 0.5$: blend fraction
when 90\% of massive halos are found, interpolated from Fig.~\ref{fig:comparem},
as well as properties not shown there:
scatter in lg(richness)-lg(mass), lg(flux)-lg(mass), and overcounting.
The latter is the number of massive halos found divided by the
total number of massive halos present.  }
\label{tab:props5}
\end{table*}
%collectstatscg,cleanfrac 0.9

\begin{table*}
\begin{center}
\begin{tabular}{|l|l|l|l|l|} \hline
\hline method &blend fraction &scatter &scatter&overcounting \\
&(finding 90\% massive) & lg(richness)-lg(mass)&lg(flux)-lg(mass)& \\ \hline
optical only & 25\% &$\geq 0.34$& -- & 15\%-29\% \\  \hline
SZ catalogue\\ + optical catalogue & 13\%&  0.19-0.20& 0.11-0.12& 16\%-20\%\\ \hline
SZ peak centered\\ optical search &15\% &0.17-0.18& 0.10 & 1\%-4\% \\ \hline
SZ peak centered\\ without merging&10\%& 0.18&0.12-0.13& 12\%-27\% \\ \hline
SZ peak centered\\ many optical &24\%& 0.17-0.34&0.10-0.34& 2\%-11\% \\ \hline
SZ alone (merged)& -- & -- & 0.10 & 17\% \\ \hline
SZ alone (unmerged)& --  & -- & 0.13    &87\% \\ \hline
\hline
\end{tabular}
\end{center}
\caption{ Properties of finders for redshift $z \sim 0.9$,
as in Table \protect\ref{tab:props5} for redshift $z\sim 0.5$.}
\label{tab:props9}
\end{table*}

Different cutoffs in richness or SZ-flux can be used in the blend fraction
measurement compared to the completeness measurement to take into
account the presence of scatter in these observables with mass.
Some of these options are discussed in the appendix of \citet{Coh07}.  
We thank the referee for suggesting we mention this point.

\subsection{Redshifts}

\begin{figure*}
\begin{center}
\resizebox{6.in}{!}{\includegraphics{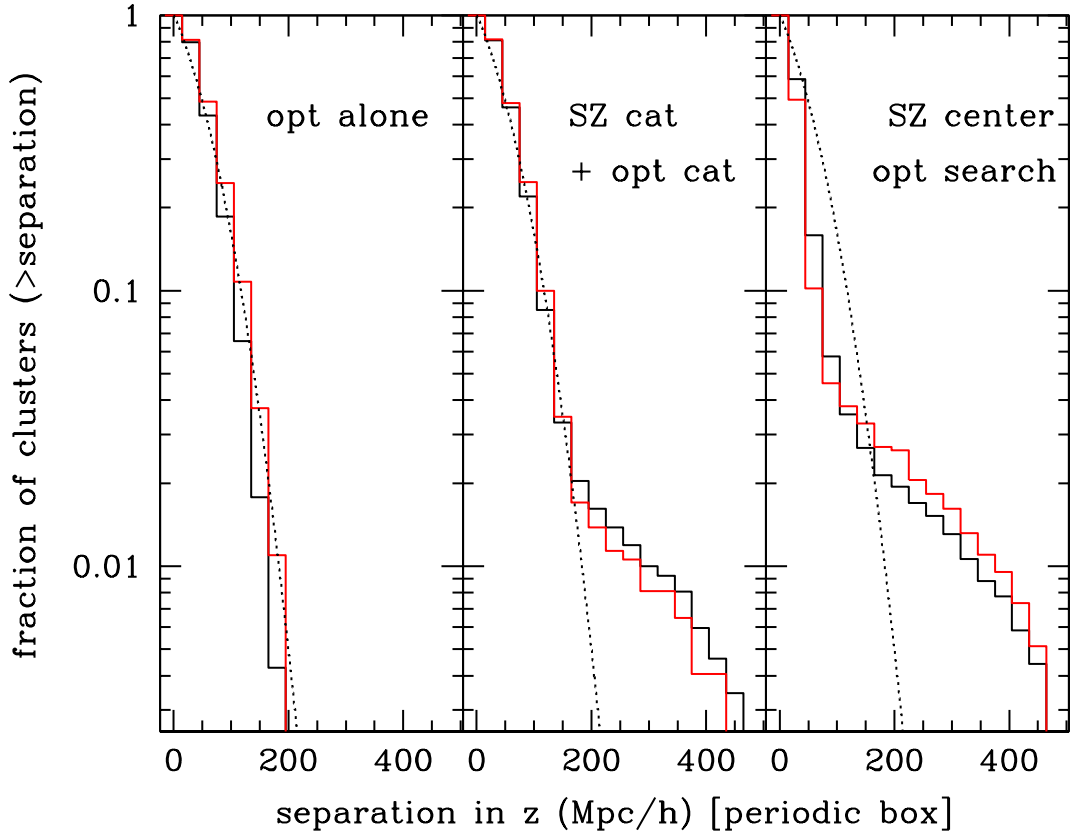}}
\end{center}
\vspace{-0.1in}
\caption{ Cumulative fraction of SZ clusters above a given
separation (measured position vs. halo true position in redshift direction)
for (left) optically found clusters, (middle) SZ clusters matched to
optical clusters, (right) optical clusters found around SZ peaks.
For the latter two, the true halo is taken to be the SZ halo (the most
massive halo in the SZ patch).
The measured position is taken to be the
redshift of the central galaxy for clusters centered on galaxies, and
the redshift slice found by the optical search for the cluster centered
on an SZ peak.  The dotted line is the cumulative fraction of objects 
with Gaussian separations corresponding to $\sim$2\% photo-$z$ errors
($\sigma=71\,h^{-1}$Mpc). The upper line at low separations for left and
central, and at high ($\sim > 150 h^{-1} Mpc$) separations for left and right, 
corresponds to halos with mass 
$\geq 2\times 10^{14} h^{-1} M_\odot$.}
\label{fig:redscat}
\end{figure*}

One of the reasons for obtaining optical information for an SZ catalogue is
to get redshifts.
The redshifts we assign are based on galaxy redshifts, which themselves have a
2\% photo-$z$ scatter in our simulations, so some scatter is expected between
the measured redshift (that of the central galaxy) and the true halo redshift
even for clusters found by optical searches alone.
The fraction of optical clusters whose redshift distance separation from
their halo exceeds a certain amount is shown in Fig.~\ref{fig:redscat},
at left.
As the box is periodic, separations are $<500\,h^{-1}$Mpc.
The scatter is similar to (but slightly smaller than) that expected
for a Gaussian separation distribution corresponding to distances
for 2\% photo-$z$ errors ($\sigma=71\,h^{-1}$Mpc), also shown.
The scatter between measured and true redshift 
is not only due to photo-$z$ scatter, since an optical cluster can match
to a different halo than that of the central galaxy, as mentioned earlier.
However, as can be seen, the actual scatter at large separations is smaller 
than the Gaussian prediction, likely because the cluster finder selects
against objects with too large separations.

The corresponding redshift errors for the SZ clusters, either those
given redshifts by finding associated clusters in an optical catalogue
(\S \ref{sec:SZopt}) or by doing optical searching around SZ peaks (\S
\ref{sec:SZcenopt}) are shown in Fig.~\ref{fig:redscat}.  For the
former case, we restrict to the 86\% of the SZ clusters whose optical
counterpart is centered within $0.5\,h^{-1}$Mpc of the peak in the
plane of the sky (as otherwise it might be expected that that the SZ
peak and optical cluster would be flagged as corresponding to
different objects).  The true redshift is taken to be that of the most
massive halo in the SZ patch (the SZ halo), the measured redshift is
either that of the central galaxy of the matched optical cluster (for
catalogue matching) or the redshift slice for the SZ peak centered
optical search.  In both, we see a tail to large separations not found
in the optical case alone.  This is due to the optical clusters whose
best matched halo is not the same as the SZ matched halo ($\sim$ 3\%
of the time for the SZ clusters matched to the most massive halos).
In addition, for the SZ centered optical searches, for separations
between 50-150 $h^{-1} Mpc$ the scatter is lower than the random scatter
expected from a 2\% photo-z scatter of the center galaxy.  In this case,
the redshift is taken to be the redshift which has the most galaxy richness
centered on it, rather than that of a specific (scattered) galaxy.  As the
galaxies are scattered independently, finding the central redshift of
the clump of galaxies is a closer approximation to the true halo redshift
than the redshift of the central galaxy.\footnote{In principle this could also
be used to find an improved redshift of the clusters that are centered on optical
galaxies.}  This leads to no improvement if the optical (albeit SZ centered) 
cluster
halo does not match the SZ halo, again leading to a large separation tail for
a small fraction of the clusters.

\section{Discussion and Summary}

Using mock galaxy catalogs and SZ flux maps made from the $z=0.5$ and 0.9
outputs of a $1\,h^{-1}$Gpc dark matter simulation, we have made a preliminary
investigation of improvements in galaxy cluster finding that arise when
optical and SZ surveys are combined.  Our errors, beam, etc. were guided by
expected properties for the DES and SPT experiments, though they are quite
similar to other experiments in progress or soon to begin.
We suggested a set of criteria by which to judge cluster finders and showed
the corresponding tradeoffs which arose in some simple cases.
Catalogues were compared using the number of massive
($\geq 2\times 10^{14}\,h^{-1}M_\odot$) halos found,
the fraction of clusters whose galaxy content was dominated by one underlying
halo, the scatter in the lg(richness)- or lg(flux)-lg(mass) relation and
the overcounting of massive halos.

As hoped, using two catalogues jointly or using optical cluster finding
centered on SZ flux peaks both led to great improvements in the catalogues.
When using the two catalogues together to catch more high mass halos, it
was preferable to take a more complete optical catalogue, 
(but more noisy, i.e. including more blends) 
and let SZ flux cuts weed out some of the outliers.  
Using SZ peaks as centers for optical searches and using 1 SZ peak per
optical cluster worked similarly in terms of blends and finding massive
halos, but worked much better in terms of overcounting.
Using a relaxed (i.e.~not merged) SZ catalogue for centering the optical
search did well for some criteria but led to a large amount of overcounting.

In the combined catalogues we have two estimates of the cluster mass.
Since the scatters in the richness- and flux-mass relations are correlated
(both arising in part from projection effects) equality of the two estimates
did not indicate a correct assignment of mass.
However, taking out $>1 \sigma$ outliers in SZ mass predictions as a function
of optical mass predictions did cut out many of the optical outliers
(i.e.~optical clusters whose predicted mass is $>2\sigma$ from the true mass).
There was a trend of SZ patches which had more optical clusters within them
to be more likely to have their (optical or SZ) predicted and true mass differ.

We also compared the redshift of the found optical cluster to that of the best
match halo to the SZ patch.  The redshift was closer if the optical cluster
was centered on the SZ peak, but even in the optical case alone there was a
scatter due to the photo-$z$ scatter of the central galaxy, and from the best
match halo not being the same as the halo of the central galaxy giving the
cluster redshift.
Except for completeness, the trends were similar at both redshifts.
At high redshift, the SZ effect was more effective at finding the massive
halos and the optical finder was worse.
Similarly, the optical finder worked better in a $\sigma_8=0.9$ catalogue
than in the ones used here, more evidence for trends of FoF vs.~SO finders
with cosmology discussed in \citet{Luk08}.

These promising results are expected, for the most part, to generalize to 
more sophisticated finders and catalogues.  Our mock catalogs were
deliberately designed to be simple, having periodic edges and no light-cone
evolution.  The catalogues underestimate the projection effects, as only
those within the box are included and do not include catastrophic photo-$z$
failures.
In addition it was assumed that the flux map could be perfectly cleaned of
foregrounds, and scatter in flux as a function of mass (due to e.g., cluster
history or hydrodynamical effects such as shocks) 
is not included\footnote{A model to incorporate some of the effects
of cluster history, especially merging, in the SZ signal in cosmological maps 
has been introduced by \citet{Bod07,ShaHolBod07}.}.

Correspondingly, our simple cluster finders did not take full advantage of the
properties of clusters or cluster galaxies, and could certainly be improved.
There are many extensions possible to the simple finders used here.
A review of optical cluster finders can be found in \citet{Gal06},
and SZ cluster finders are improving as well
\citep[e.g.][]{SchWhi03,DelMelBar02,GeiKneHob05,PieAnt05,Pie05,Pir05,
Sch06a,Sch06b,MelBarDel06}.
In particular, using profile information
\citep[such as used in matched filters, e.g.][]{Pos96,WhiKoc02,Koc03,Don07},
changing the way halos are assigned to SZ peaks, using the peak size,
etc., can all refine the cluster information.
One extreme, if one is only interested in cluster finding as a route to
cosmological parameters, would be to bypass the cluster finding entirely
and just cross-correlate the SZ flux and galaxy maps, and compare to mock
catalog predictions for the cross-correlation.

Several issues we raised for the simple finders will become more pressing
with more sophisticated finders -- cluster finders with increased complexity
depend more and more upon the expected properties of the targets.
In particular, finders can succeed or fail in several different ways and
choices must be made about which failures/successes one cares most about.
Specifically, as finders will be increasingly tuned and trained on mock
catalogues, and no finder will be perfect, the question arises as to which
errors are preferable.
Ideally the errors should not only be as small as possible, but also well
modeled in the mock catalogues.
It is preferable to have a finder which relies most heavily upon the most
accurate features of the mock catalogues at hand.
This is especially important as the finders become more complex and assumed
cluster properties become more and more implicit.
In addition, there is a question of how the scatter in the finders depends
on the target cluster masses, for example higher mass lowers the importance
of shot noise errors on the optical richness and is more likely to correspond
to a significant SZ signal.  Our focus was on massive
($\geq 2\times 10^{14}\,h^{-1}M_\odot$)
halos, as the SZ methods cannot probe to very low masses, but the particular
science goals may argue for a different threshold.
The tradeoffs chosen in creating the finder should in part be decided by
the intended use of the sample.
Usually, the more complex the finder becomes the more tradeoffs are necessary
in desirable catalogue features.
These decisions are best made with an eye the strengths, features and science
goals of the particular observational data set in hand, and the strengths of
the mock catalogues available to analyze the data. 
As such multi-method cluster observations increase,
the exciting science within reach will also increase dramatically.

JDC thanks C. Heymans, J. Kollmeier, A. Leathaud, A. Meiksin, R. Nichol, 
P. Norberg, W. Percival, C. Pfrommer and E. Rozo for discussions.  We would
like to thank U.~Portsmouth, the ROE and IUCAA for hospitality and the
opportunity to present this work while in progress and completed.  We thank
G. Evrard for comments on the draft and the anonymous referee for very helpful
comments on the submitted paper.
JDC acknowledges NSF-AST-0810820 and DOE for additional support.
MW is supported by NASA and the DOE.
The simulations used in this paper were performed and analyzed at the
National Energy Research Scientific Computing Center.

\citet{Men08} appeared when we were preparing this paper for publication, 
which discusses
optical clusters found in the Southern Cosmology Survey and their
expected SZ signal from ACT.

\end{document}